\documentclass[a4paper,10pt]{article}

\usepackage{hyperref}
\usepackage{dcolumn}
\usepackage{a4wide}
\usepackage{graphicx,setspace}
\usepackage{bm}
\usepackage{amsmath,textcomp}
\usepackage{amssymb}
\usepackage{subfig}
\usepackage{xcolor}
\usepackage{booktabs}

\usepackage{graphicx,setspace}
\usepackage{bm}
\usepackage{amsmath,textcomp}

\usepackage{romannum}
\usepackage{xcolor} 
\usepackage{changes}
\definecolor{darkgray}{rgb}{0.66, 0.66, 0.66}
\definecolor{darkviolet}{rgb}{0.58, 0.0, 0.83}
\definechangesauthor[name={Lock}, color=darkviolet]{LY}
\definecolor{pakistangreen}{rgb}{0.0, 0.4, 0.0}
\definechangesauthor[name={Ning}, color=pakistangreen]{NN}
\definecolor{redred}{rgb}{0.99, 0.0, 0.0}
\definechangesauthor[name={Stefan}, color=redred]{ST}

\newcommand{\blue}{\color{black}}

\begin{document}
\vspace*{-3cm}
\begin{flushleft}
{\Large
\textbf{Bali's ancient rice terraces: A Hamiltonian approach}
}
\\
Y\'{e}rali Gandica$^{1,\ast}$,
J. Stephen Lansing$^{2,3}$ and
Ning Ning Chung$^{4}$
Stefan Thurner$^{5,6}$
Lock Yue Chew$^{7,8}$
\\
\singlespacing
\footnotesize {
{1}  CY Cergy Paris Université, CNRS, Laboratoire De Physique Théorique et Modelisation, F-95000 Cergy, France\\
{2} Santa Fe Institute, 1399 Hyde Park Road, Santa Fe, NM 87501, USA. \\
{3} Stockholm Resilience Centre, Kraftriket 2B, 10691 Stockholm, Sweden. \\
{4} Centre for University Core, Singapore University of Social Sciences, Singapore 599494. \\
{5} Section for Science of Complex Systems, Medical University of Vienna, Spitalgasse 23, A-1090 Vienna, Austria.\\
{6}Complexity Science Hub Vienna, Josefst\"adterstra\ss e 39, A-1080 Vienna, Austria.\\
{7} School of Physical \& Mathematical Sciences, Nanyang Technological University, Singapore 637371.\\
{8} Data Science \& Artificial Intelligence Research Centre, Nanyang Technological University, Singapore 639798.
}
\end{flushleft}

\singlespacing

\begin{abstract}
We propose a Hamiltonian approach to reproduce the relevant elements of the centuries-old Subak irrigation system in Bali, showing a cluster-size distribution of rice-field patches that is a power-law with an exponent of $\sim2$. Besides this exponent, the resulting system presents two equilibria. The first originates from a balance between energy and entropy contributions. The second arises from the specific energy contribution through a local Potts-type interaction in combination with a long-range anti-ferromagnetic interaction without attenuation. Finite-size scaling analysis shows that as a result of the second equilibrium, the critical transition balancing energy and entropy contributions at the Potts (local ferromagnetic) regime is absorbed by the transition driven by the global-antiferromagnetic interactions, as the system size increases. 
The phase transition balancing energy and entropy contributions at the global-antiferromagnetic regime also shows signs of criticality. Our study extends the Hamiltonian framework to a new domain of  coupled human-environmental interactions.
\end{abstract}



The delicate balance between energetic and entropic contributions is responsible for those singular points in parameter space where phase transitions occur. The nature of these transitions is well studied in physics. However, phase transitions are more and more recognised in complex systems, where interactions can have a much richer structure \cite{Thurner2018,Sornette2000}. Typical systems in physics consider interactions of either short- or long-range. Criticality has been found for both cases \cite{Stanleybook,Goldenfeld1992,Bayong1999,Flores_Sola_2015}. Critical transitions are characterized by the divergence of the correlation length, which leads to simplifications of various thermodynamic functions; these often take the form of power-laws \cite{Christensen2005}. 

It is intuitive to define energy in physical systems, but how can this be generalized to other systems? Motivated by the power-law size-distributions of irrigated rice terraces that are managed by farmer associations called {\em Subaks} in Bali-Indonesia \cite{Lansing2017}, in this letter  we present a Hamiltonian formulation aimed to represent the  most relevant interactions in managing Balinese rice paddies, without being distracted by myriad confusing details \cite{Buchanan2002}.

Since the 11th century, Balinese farmers growing paddy rice had to balance two opposing constraints. On one side they are confronted with shortages of irrigation water, and on the other they have to control rice pests by synchronized flooding of fields (which reduces the habitat of the pests). 
These constraints are represented in our proposed Hamiltonian by two types of interaction, one being short- the other long-range. 
Because most rice pests can move, synchronizing irrigation and planting schedules is essential so that fields become fallow at the same time deprives pests of their food. In the Hamiltonian this is captured by incentivizing neighbouring cells to be in the same planting-state, meaning that farmers follow the same harvest schedule. The second ingredient represents the fact that water is a limited resource, resulting in a global constraint. The larger the area that follows an identical irrigation schedule to control pests, the more peak irrigation demands will coincide, which reduces the amount of water available for neighboring farms. The Hamiltonian has a corresponding long-range interaction term that favours disorder. The two constraints have opposing effects: the larger the agricultural area that follows the same irrigation schedule, the more water stress appears from the synchronized irrigation cycles. The irrigation schedule in each field can be in one of two states with respect to its neighboring fields: synchronized or unsynchronized. The Ising with two and the Potts model with $q$ states are both prototypes for discrete spin systems. In the Potts model, for a number of $q>4$ states, the phase transition is first-order and becomes critical or second-order for $q\leq4$  \cite{Wu1982}.
The scenario for $q\approx 4$ remains interesting in statistical mechanics. While the corresponding  correlation lengths do not diverge, they are still very high and several thermodynamic functions can be represented as scaling forms. These so-called weak first-order transitions, are hard to distinguish from second-order; both show strong correlations \cite{Peczak1989,Gandica2016}. In both cases, the specific geometry of the interaction geometry specifies the propagation of fluctuations through the system.

\begin{figure}[t]
\begin{center}
\includegraphics[width=0.7\columnwidth]{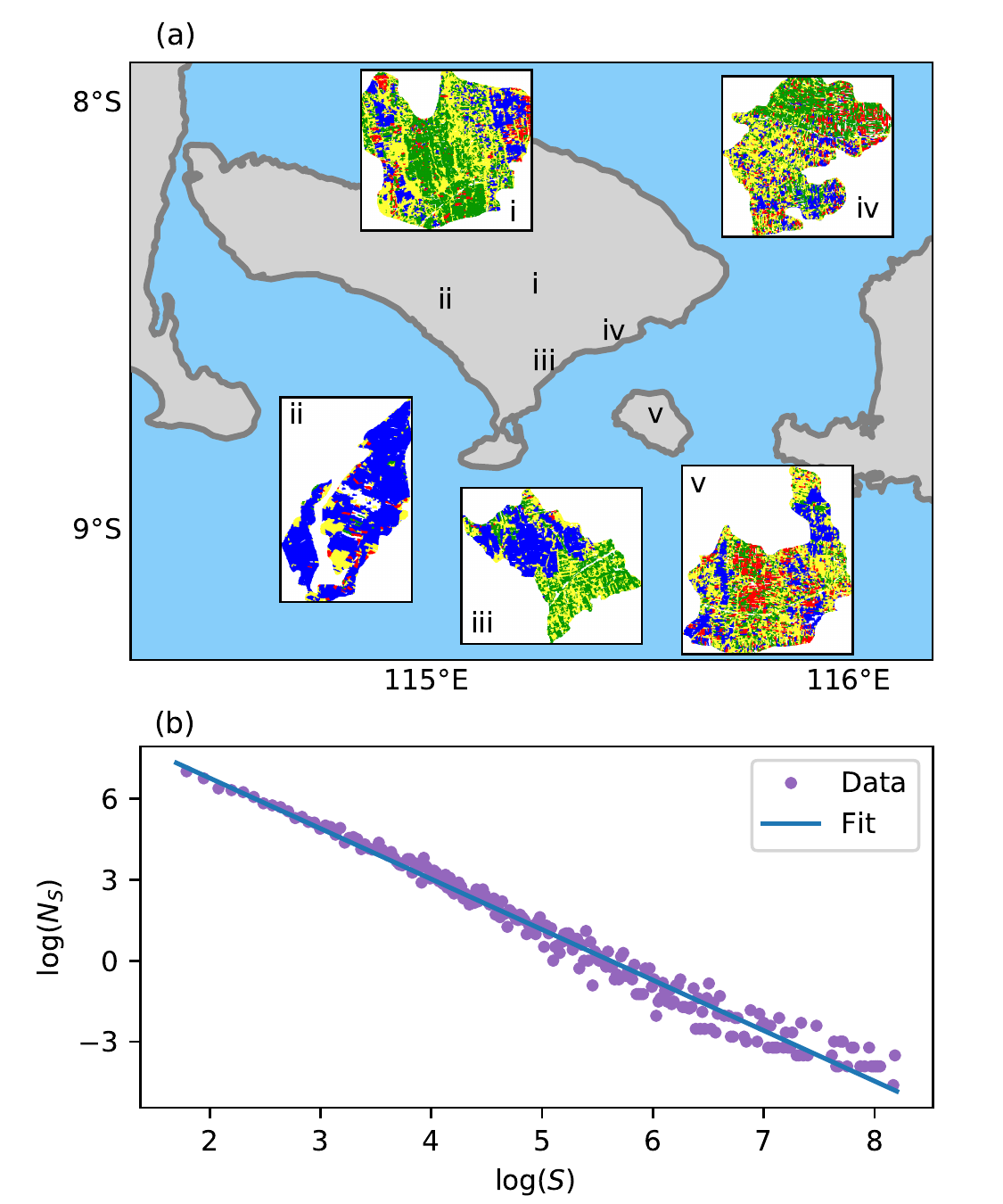}
\end{center}
\vspace{-0.5cm}
\caption{
(a) The 5 regions where the data was taken: i. Gianyar, ii. Tabanan, iii. Sukawati, iv. Kusamba and v. Klungkung.
(b) Log-log plot for the cluster-size distribution of the aggregated  normalised values of all the regions and different years, where photosynthetic activity was measured (13 in total). The corresponding exponent is $\tau \sim 1.87(2)$. 
}
\label{fig1}
\end{figure}

{\blue In \cite{Lansing2017}, the growth and harvest cycle of Balinese rice fields was divided into four stages: grow, harvest, flood, drain.} 
Every field is classified to be in one of these four states by estimating the photosynthetic activity by  multi-spectral and panchromatic satellite images \cite{Lansing2017}. 
{\blue The value of the power-exponent of the cluster-size distribution (the Fisher exponent) was reported as about $\tau\sim2$.
}
This inspired the authors to propose an adaptive, self-organized process that explains the emergence of power-law distributions in {\blue coupled human–natural systems.} Balinese have grown paddy rice for at least one thousand years, making it plausible that the many Balinese irrigation systems had ample time to adapt to a globally  optimal (or close to optimal) situation in each region. 
Figure \ref{fig1}-a shows the 5 regions in Bali, where the data were taken.
The multi-spectral analysis of each individual regional sub-system (visible in Fig. \ref{fig1}-a) in \cite{Lansing2017} presents a power-law distribution with similar exponents ranging between $1.76$ and $2.19$.

In Fig. \ref{fig1}-b, we show the cluster-size distribution of the data on Balinese rice fields, as used in \cite{Lansing2017}, with the slight difference that here we aggregate normalised values of all 5 regions and different years when photosynthetic activity was captured (13 in total). 
{\blue Here, a cluster is defined as a connected region of sites that are in the same state of cropping activity. To determine a cluster, we use the breadth-first search algorithm that searches for connections among neighboring sites in the same state. The size of a cluster is then defined as the total number of sites in the cluster.
To define a site note that a rice farm in a Bali spans typically one-third of a hectare ($\approx 3330m^2$). The analyzed satellite imagery covering a Subak typically contains $700 \times 700$ pixels, one pixel that represents a site here is equivalent to an area of about 25$m^2$. The analyzed image thus corresponds to a lattice of approximate dimensions of 60 farms $\times$ 60 farms. 
The exponent of the cluster-size distribution is found to be $\tau=1.87 (2)$, in line with the previously reported values.} 

We start with a Subak Hamiltonian defined as: 
  \begin{equation}
   H_S = -a \sum_{\langle i,j \rangle} \delta (\sigma_i,\sigma_j) + \frac{b}{N-1} \sum_{i>j} \delta (\sigma_i,\sigma_j) \, ,
  \end{equation}
where $\sigma_i$ represents one of the $q$ states at site $i$. Since the fields can be in four states, throughout the paper, we set $q=4$. $\langle i,j \rangle$ means the sum over four neighboring sites, $\delta (\sigma_i,\sigma_j)$ is the Kronecker delta. The  second sum extends over all pairs on a $2D$ square lattice of linear size $L$. $N$ is the total number of sites on the lattice, i.e. $N = L \times L$.
$a$ represents the level of pest stress, defining the local interaction between nearest neighbours. Pest stress acts as a ferromagnetic Potts model, promoting ordering at low temperatures. The Hamiltonian system balances this local interaction with a long-range anti-ferromagnetic contribution. The effect of  limited water supply is regulated by $b$, and is a global or system-wide interaction. This contribution has the shape of a mean-field Potts model with a positive sign and drives the system towards states where the $q$ states appear with approximately the same frequency. Note that this global interaction does not have a distance attenuation factor, as often used for long-range interactions \cite{Flores_Sola_2015}; it has a factor, $1/(N-1)$, to balance the local contribution.

The total energy, $\epsilon$, of the Subak system is a fixed quantity defining the condition of equilibrium in a canonical formulation \cite{Lee2012}. The maximum entropy principle finds the most likely distribution function for the allowed configurations consistent with a fixed $\epsilon$. Fixing the expectation value of energy to $\epsilon$ and normalizing with Lagrangian multipliers, by assuming processes being reasonably close to i.i.d., functional variation of the Gibbs-Shannon entropy yields 
\begin{equation}
 \delta \left[\sum_{i=1}^N p_i \ln p_i - \beta \left(\sum_{i=1}^N p_{i} {\epsilon}_i -\epsilon \right) + \nu \left( \sum_{i=1}^N p_i - 1 \right) \right] = 0 \,.
\end{equation}  
the Boltzmann distribution
\begin{flalign}
 p_i=\frac{e^{-\beta \epsilon_i}}   { \sum_{i=1}^N e^{-\beta \epsilon_i}} \,,
\end{flalign}
where inverse temperature, $\beta$, that controls the fluctuations in the dynamics, is fixed by the average energy 
{\blue
\begin{flalign}
\frac {\sum_{i=1}^N \epsilon_i e^{-\beta \epsilon_i} }{ \sum_{j=1}^N e^{-\beta \epsilon_j} } = \epsilon_P + \alpha \,.
\end{flalign}
}
In this way, we can think of a fixed temperature for the ensembles, or of the average energy (Potts energy $\epsilon_P$ + a constant value $\alpha$). 
{\blue Obviously, ``temperature'' can be associated with ``non-rational'' decisions on the side of Subaks for both interaction types. Without temperature, any dynamics would lead to local minima. On the technical side, temperature (irrational moves from time to time) helps us to approach global energy minima (global maximum of entropy).
}

We implemented the model with a Metropolis Importance Sampling Monte Carlo (MC) method for the simulations. We consider both, open and periodic boundary conditions. Note that, contrary to open boundary conditions, periodic boundary conditions lead to additional contributions in the first term of Eq. 1, shifting the critical temperature towards higher temperatures, when $b$ is small. This impact diminishes with larger system sizes and becomes negligible when $b$ gets large with respect to $a$. {\blue Results for both boundary conditions are very similar; we only show those for the open ones. Only the figure for the collapse (Fig. \ref{fig6}) has been done using periodic boundary conditions to better simulate the thermodynamic limit ($N \rightarrow \infty $).
We use $L^2*10*q$ MC thermalization steps to avoid transients and $L^2*100*q$  MC steps for the averages (except for Fig.~\ref{fig6} where we used $L^2*500*q$ MC for both the thermalization and also for the averages). We show results for different system-sizes.}

\begin{figure}[tb]
\begin{center}
\includegraphics[width=0.6\columnwidth]{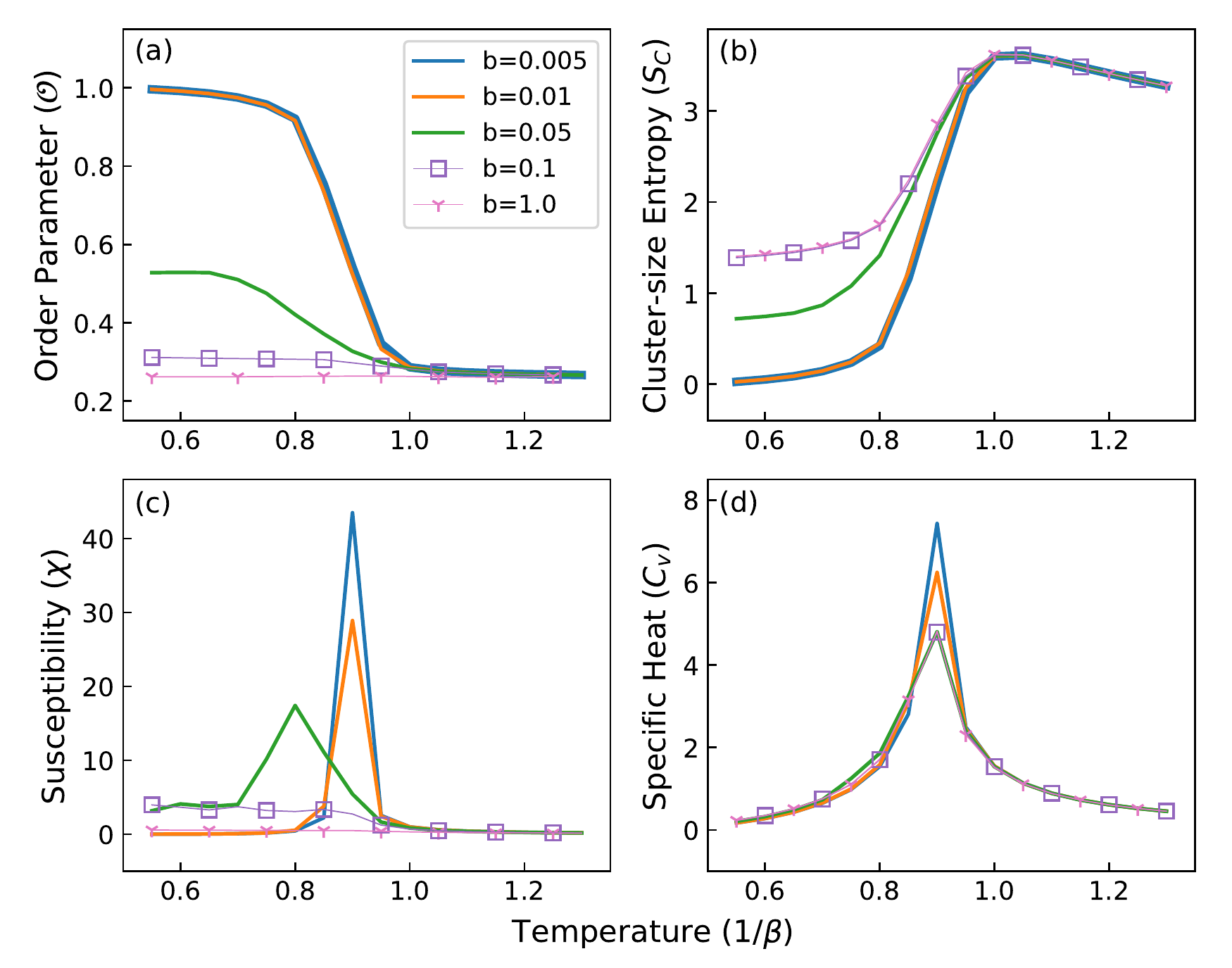}
\end{center}
\vspace{-0.5cm}
\caption{(a)  Order parameter, (b) Cluster-size entropy, (c) Susceptibility and (d) Specific heat, for the Subak Hamiltonian on grids of linear size, {\blue$L=60$ with open boundary conditions}. Pest stress is fixed to $a=1.0$. We show several values of water stress, $b$. {\blue Color code in (b)-(d) is the same as in (a).}
}
\label{fig2}
\end{figure}

\begin{figure}[t]
\begin{center}
\includegraphics[width=0.6\columnwidth]{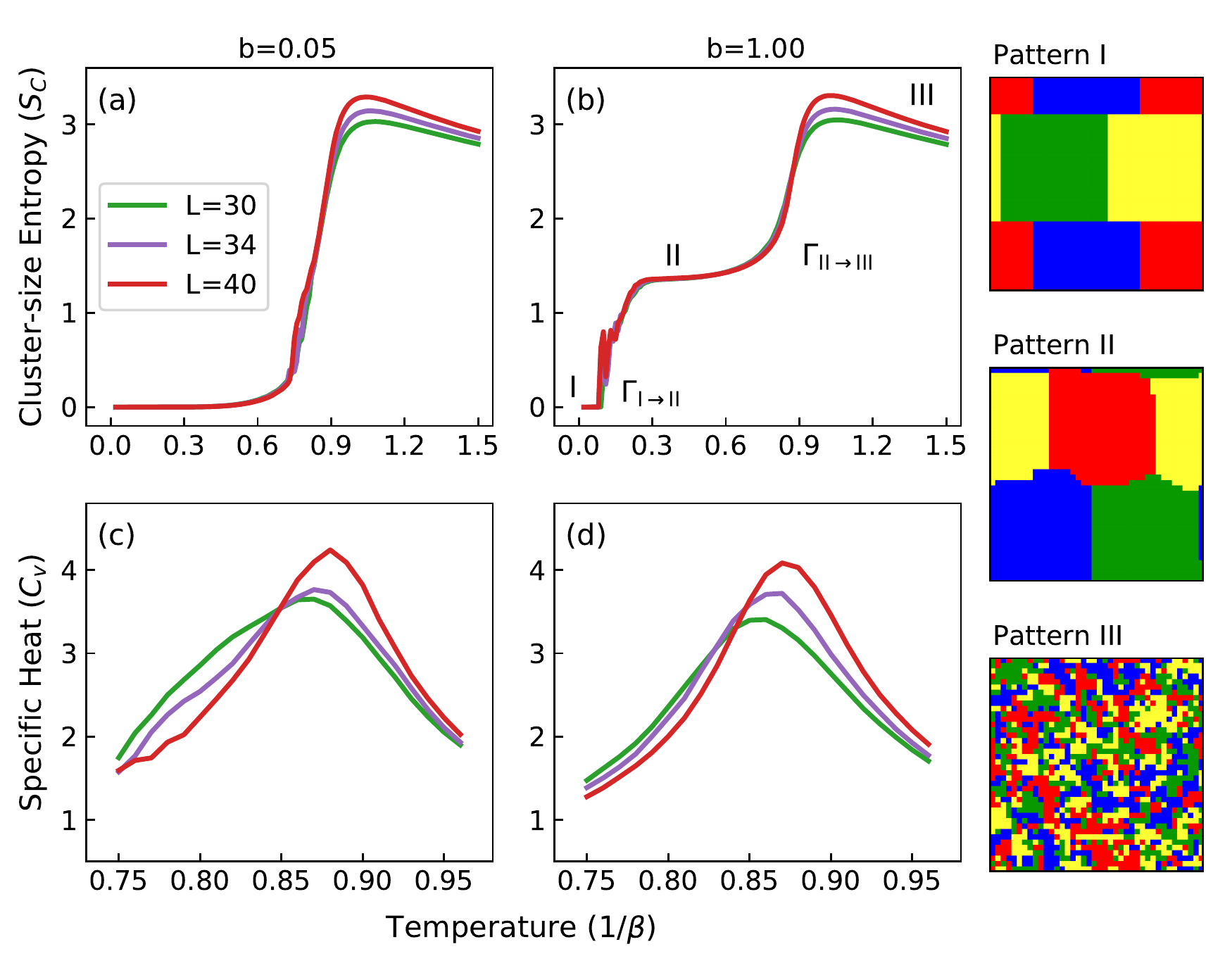}
\end{center}
\vspace{-0.5cm}
\caption{Cluster-size entropy (a) and (b) and specific heat (c) and (d) for the Subak Hamiltonian with $a =1.0$ and two different values of $b$. Colors indicate system size $L=30,34$, and $40$. For $b=1$,
$T_c$ moves (peak of specific heat) to high temperatures with increasing system size. 
For $b = 0.05$, $T_c$ moves towards lower temperatures with increasing size. 
{\blue Color code in (b)-(d) is the same as in (a).}
}
\label{fig3}
\end{figure}

\begin{figure}[t]
\begin{center}
\includegraphics[width=0.6\columnwidth]{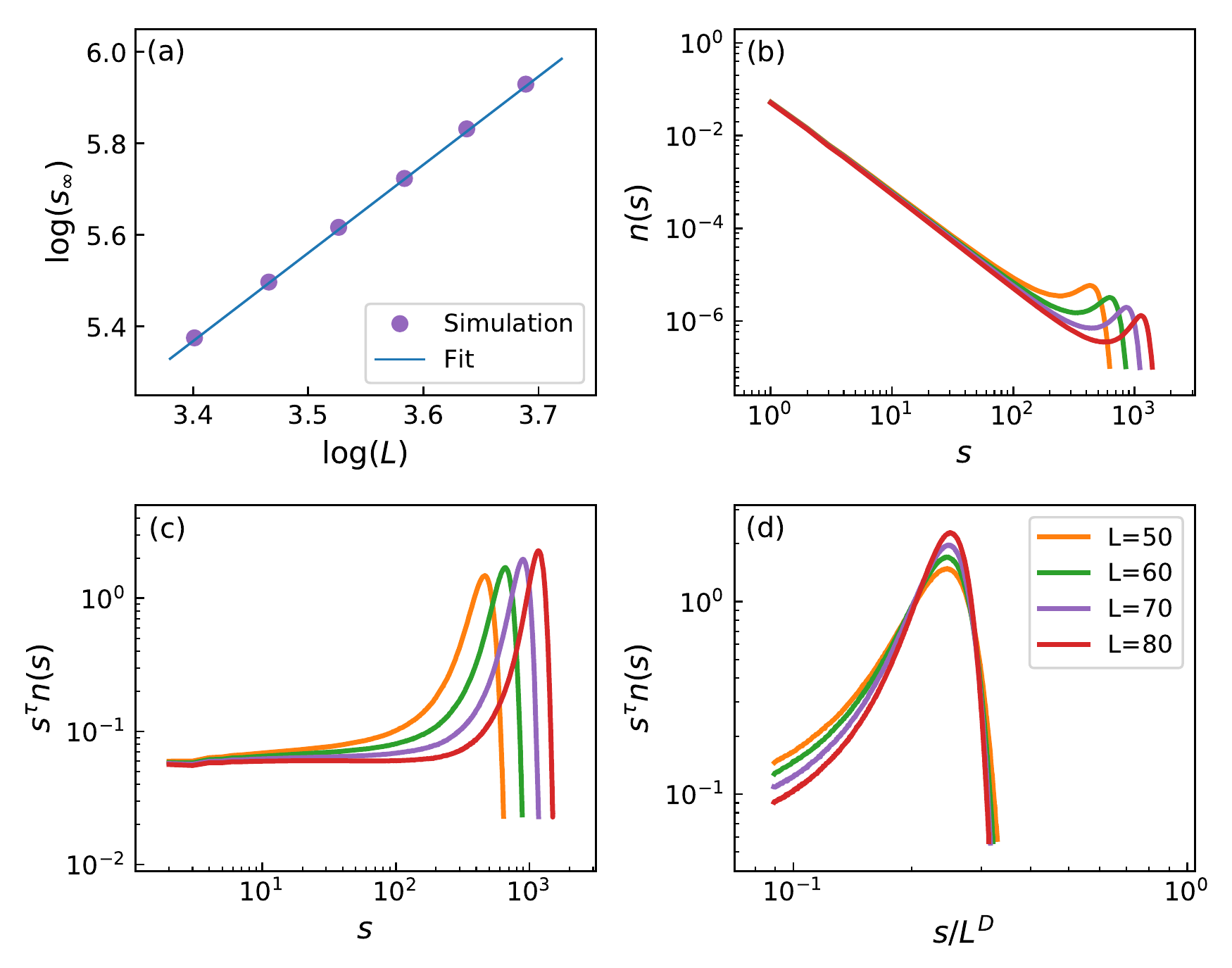}
\end{center}
\vspace{-0.5cm}
\caption{ Fractal dimension derived from the way the percolating cluster fills the space (a). We find a slope of $D=1.927(4)$. In (b)-(d) we show an attempt to collapse the cluster-size distribution onto a single function for various values of $L$. We find a typical scenario for a weak first-order transition. 
{\blue Color code in (b)-(c) is the same as in (d).}
}
\label{fig6}
\end{figure}

Figure \ref{fig2}-a depicts the order parameter, $\cal{O}$, that is defined as the  fraction of the largest cluster in any state. Figure  \ref{fig2}-b shows the cluster-size entropy, as defined in \cite{Tsang1999}, which measures the entropy for the clusters of different sizes in the system,
{\blue
\begin{equation}
 S_C = - \sum_s P_s \, \ln  P_s  \,,
\end{equation}
}
where $P_s$ is the probability that a site belongs to a cluster of size, $s$.  $S_C$ has been shown to succeed in locating transition points \cite{Gandica2011}. We use it as a suitable function to indicate the transition for large values of the long-range contribution, $( b \geq 0.1)$, that is not visible by the order parameter, see also Fig. \ref{fig3}-b. 
{\blue We show two response functions, the susceptibility defined as
\begin{equation}
    \chi = \frac{\langle {\cal O}^2 \rangle - \langle {\cal O} \rangle^2}{T} \,,
\end{equation}
in Fig. \ref{fig2}-c, and the specific heat, 
\begin{equation}
    C_v = \frac{\langle H^2 \rangle - \langle H \rangle^2}{NT^2} \, ,
\end{equation}
is seen in Fig. \ref{fig2}-d.} 
We show these results for different values of the global contribution weight, $b$, after fixing the intensity of the local interaction to $a=1$. 

The Potts regime is visible in Fig. \ref{fig2} for small values of the water stress parameter, $b <0.05$. Large $b$ values ($b \geq 0.5)$ do not show a change in the order parameter across temperature (see Fig. \ref{fig2}-a). However, the transition is visible on the cluster size distribution shown in Fig. \ref{fig2}-b. The low temperature regime is ordered (i.e. all sites are in one state) at low $b$, as expected for the Potts regime. {\blue The  minimum  energy  state  is  size-dependent  because  of  the interplay between the local and global contribution.} As $b$ increases, the system splits into two approximately equally sized clusters of two states, finally, for high $b$, reaching a $q$-balanced state, where the system is mainly driven by the global anti-ferromagnetic interaction. 
From the peaks of the susceptibility and the specific heat, shown in Figs. \ref{fig2}-c and \ref{fig2}-d respectively, we see that the order-disorder transition points move to lower temperatures, as $b$ increases, before reaching a fixed point. The anti-ferromagnetic contribution reinforces the entropic effects, favoring disorder and the subsequent reduction of critical temperatures. 

We now continue with to analyse the short-range Potts-like transition by first fixing $b = 0.05$, and then the transition driven by the global contributions with setting $b = 1.0$. Figure \ref{fig3} shows the cluster-size entropy in (a) and (b) for the two respective $b$ values, and the specific heat (c) and (d). There, two completely different scenarios become visible. Note that there are two contributions to the energy: the total boundary between the coloured patches (same Potts energy) and the one that is related to the balance of frequencies of the $q$ states. 

At $b=1$, where local and global contributions are balanced, a plateau in the cluster-size entropy, see Fig.\ref{fig3}-b, indicates two structural transitions. The low-temperature transition, indicated by $\Gamma_{\mbox{\Romannum{1}} \rightarrow \mbox{\Romannum{2}}}$, is the result of a competition between maintaining a short (straight) boundary (local interaction) between patches and developing curvature at the boundary (global interaction). The dominance of the former leads to four equal-sized patches (${\cal O} = 0.25$, $S_C=0$), while a balance between the former and the latter causes four slightly unequal-size patches (${\cal O} = 0.25$, $S_C=\ln(4)=1.386$). This balance is reflected in a robust plateau in ${\cal O}$ over an extended temperature range before it is disrupted by entropic effects in a order-disorder transition, indicated by $\Gamma_{\mbox{\Romannum{2}} \rightarrow \mbox{\Romannum{3}}}$. The Potts regime ($b=0.05)$ is different. There, the system remains in one state until the regular order-disorder transition occurs, Fig. \ref{fig3}-a.

Note that for $b=1$, the energy contribution balances the frequencies of the $q$ states and does neither promote order nor disorder. The system is already balanced at low temperatures and continues like that also at higher ones. As a consequence, the energy contribution does not change that transition at high temperatures. Instead, this  transition is the result of the entropic contribution. 
We see in Fig. \ref{fig3}-d that $T_c$ (peak of specific heat) moves towards higher temperatures with increasing system size, $L$.
To understand finite-size effects, note that the global anti-ferromagnetic contribution diverges, as $L \rightarrow \infty$. The lack of that strong energy contribution in finite systems is responsible for the appearance of the higher critical temperatures as $L \rightarrow \infty$.   

Another consequence of the increasing influence of the global anti-ferromagnetic term with increasing system size, is that it dominates the system in the thermodynamic limit, and destroys the apparent local-ferromagnetic Potts-like transition at low $b$. This is why  the height of the specific heat does no longer increase systematically with system size for $b = 0.05$, see Fig. \ref{fig3}-c. 
Keeping in mind that the apparent transition dominated by the local contributions disappears in the thermodynamic limit, we focus on the global anti-ferromagnetic-driven transition. {\blue To find the critical temperature, we use the scaling Ansatz, $T_c(L)=T_c(\infty)+ \lambda L ^{-1/\nu}$, and get $T_c(\infty)=0.89(2)$.} The calculation of the critical exponents is beyond the scope of this work.

In Fig. \ref{fig6}-a, we show the calculation of the fractal dimension, measuring how the percolating cluster fills the space, $S_\infty \propto L^D$. We find $D=1.927(4)$ and the exponent to the cluster-size distribution, $\tau=2.036(1)$.
Here we used the famous relation $(\tau=\frac{d}{D}+1)$, where $d=2$ is the dimension. Figures \ref{fig6}-b to (d) show the data collapse, using the finite-size scaling Ansatz for the cluster number density at the critical point \cite{Christensen2005}
\begin{equation}\label{eq:scaclus}
n(s,T_c;L) \propto s^{- \tau} \phi(s/L^D ),\quad L\gg 1, s\gg 1 \, ,
\end{equation} 
to determine the nature of the high-temperature transition. Figure \ref{fig6}-d shows a resemblance collapse. Although the collapse is not perfect, the system clearly shows signs of criticality. 
{\blue The transition is ––if not second order–– at least weak first order. It is important to note that the approximate data collapse occurs as a consequence of the divergence of the correlation length. 
}

We proposed a Hamiltonian system based on the main mechanisms behind farmer's planting decisions. Our interest focused on understanding the effect of a Hamiltonian approach that balances local and global contributions. In summary, the system described by the Subak Hamiltonian shows a highly non-trivial structure, in particular it exhibits two types of equilibria. The first arises from a balance between the energy and entropy contributions, as known for conventional Hamiltonians. The second equilibrium comes from a balance in the local and non-local energy contributions. A finite-size scaling analysis shows that as a result of the second balancing mechanism, the transition from energy and entropy contributions can get absorbed by the transition for high values of the global anti-ferromagnetic contribution, as system size increases. The system thus seems to present a weak-first order phase transition in the thermodynamic limit. The discovery of weak first-order transitions is consistent with the strong correlations in the system, as we would expect them to be present in reality. 

{\blue We conclude that the model Subak system self-organises towards a situation of a  balanced equilibrium (at the critical temperature), as a consequence of the strong correlations between the farmers' planting schedules, which gives rise to the power-law of the cluster-size distribution.} 

{\blue The exponent obtained from the Hamiltonian system, $2.036$, is reasonably close to the one obtained in the game theoretical framework \cite{Lansing2017},  $1.9$, and clearly falls within the range of the empirical estimates from the observed patch-size distribution in Balinese Subaks.
}
Our study focused on $q=4$ states, {\blue and extends the game theoretical framework with entropic effects in a thermodynamic setting by including a competition between energy and entropy that gives rise to an order-disorder transition with signs of criticality.} Future analysis for different $q$ values would be a natural next step. Having access to patch-size distributions for different $q$ values and comparing the corresponding cluster-size distributions with the predictions from the Hamiltonian approach, could provide interesting insights on the soundness of a thermodynamic limit ($L \rightarrow \infty$), in critical social or social-ecological systems. 

We close by addressing the question of possible consequences of pushing systems away from their adaptive equilibrium. An example is the effects of the introduction of  Green Revolution agriculture to Bali in the 1970s. At that time, the Subaks were required to give up the right to set their irrigation schedules. Instead, each farmer was instructed to cultivate Green Revolution rice as often as possible, resulting in unsynchronised planting schedules. By 1977, 70\% of southern Balinese rice terraces were planted with Green Revolution rice. At the beginning, rice harvests increased. However, within 2 years, Balinese agricultural and irrigation workers reported the catastrophic ``chaos in water scheduling'' and ``explosions of pest populations'', namely, the triggering of a system-wide catastrophe \cite{Lansing2006}.

\section*{Acknowledgements}
Computational resources were provided by  
Consortium des Équipements de Calcul Intensif (CÉCI), 
cluster Osaka of the Le Centre De Calcul (CDC) of the Direction Informatique et des Systèmes d’Information (DISI) de l’Université de Cergy-Pontoise, 
and the School of Physical \& Mathematical Sciences of the Nanyang Technological University (NTU) in Singapore.
We acknowledge funding from the Fonds de la Recherche Scientifique de Belgique (F.R.S.-FNRS) under Grant No. 2.5020.11 and by the Walloon Region. 
YG thanks the Visiting Fellowship provided by the Complexity Institute at NTU and 
thanks Ismardo Bonalde, Silvia Chiacchiera, Bertrand Berche, Petter Holme, and Ernesto Medina for helpful discussions. All authors thank Janusz A. Holyst and Sydney Redner for stimulating discussions.

\end{document}